\documentclass[conference]{IEEEtran}
\usepackage{cite}
\IEEEoverridecommandlockouts 
\usepackage{balance}
\usepackage[pdftex]{graphicx}
\usepackage{caption}
\usepackage{subcaption}
\usepackage{amsmath,stmaryrd,pict2e,picture}
\usepackage{float}
\usepackage{amsmath,amssymb,amsthm}
\usepackage{mathrsfs}
\usepackage{algorithm}
\usepackage{algpseudocode}
\usepackage{color, soul}
\usepackage{multirow}
\usepackage{graphicx}

\begin{document}
\title{Deep Learning-Based Sparse Array Design with Pre-Steering for Adaptive Beamforming}
\author{
\authorblockN{Ian~Straub\authorrefmark{1} and Syed~A.~Hamza\authorrefmark{2}}
\authorblockA{School of Engineering, Widener University, PA, USA \\
Emails: \authorrefmark{1} ikstraub@widener.edu, \authorrefmark{2} shamza@widener.edu}
\thanks{This work is funded by the National Science Foundation (NSF) Award \#2347220.}
}

\maketitle
\begin{abstract}
This paper investigates the use of convolutional neural networks (CNNs) for learning sparse array configurations that achieve near-optimal beamforming under varying source and interference angles. Unlike conventional or convex optimization based algorithms, the proposed deep learning approach enables rapid reconfiguration of sparse arrays in highly dynamic propagation environments. The paper considers a single desired source and a single interference signal at arbitrary angles, analyzing scenarios with both fixed and varying desired source directions. To avoid retraining for each possible source angle, an array pre-steering strategy is introduced, whereby the network is trained only at broadside, while test inputs are pre-steered to align with the broadside direction. To account for practical imperfections, the effect of pre-steering errors is examined, and a robust error-augmented training is adopted. The approach systematically incorporates small, structured pre-steering perturbations during training, enabling the network to maintain high classification accuracy and maximize the signal-to-interference-plus-noise ratio (SINR) even under angular uncertainty. The results demonstrate that the proposed method achieves over 90$\%$ test accuracy across wide ranges of source and interference angles, highlighting its potential for real-time, robust sparse array configuration in dynamic environments.

\end{abstract}

\section{Introduction}
Sparse arrays have long been utilized in radar, sonar, and wireless communication systems due to their ability to achieve high spatial resolution and interference suppression with a reduced number of active antennas \cite{ref:haykin1980array, AM68, PV10, QZA15, AWZ16, RK21, ref:amin_sparse_arrays_2023}. By exploiting non-uniform sensor placement, sparse arrays provide additional degrees of freedom for beamforming while maintaining low hardware complexity and cost. Conventional optimization-based methods, including iterative and semidefinite relaxation (SDR)  algorithms \cite{ref:zheng_sig_proc_letters_2020, ref:wen_vehicular_tech_2021, HPP12, WAT14, GVK18, ref:boudaher_dsp_2020, ref:hamza_sig_processing_2019, 8378759, 8682266, mypaper, 9257073, 9114655, HAMZA2020102678} have been developed to design sparse array configurations that achieve maximum signal-to-interference-plus-noise ratio (MaxSINR) beamforming. Despite their effectiveness, these iterative techniques are computationally demanding and often unsuitable for real-time operation in rapidly changing propagation environments.

Recent advances in deep learning (DL) \cite{ref:lecun_dl_2015, ref:goodfellow_dl_2016, GU2018354} have enabled data-driven approaches to sparse array beamforming and wireless communications \cite{ref:li_wcmc, DBLP:journals/twc/ShenSZL20, 8444648, DBLP:journals/corr/OSheaEC17}, including DL-based direction-of-arrival (DOA) estimation and antenna selection \cite{ref:wandale_dl_doa_2021, icassp23, 9266359}.  Once trained, a neural network can quickly infer the optimum array configuration from the input covariance data, facilitating rapid adaptation to dynamic conditions. In prior work, convolutional neural networks (CNNs) were trained to predict optimum sparse array configurations \cite{11031846, doi:https://doi.org/10.1002/9781394191048.ch7, 10149410}, enabling MaxSINR beamforming from data-driven models.  To generalize across source directions, training datasets included all possible desired angles, which, although effective, required very large datasets and substantial computational resources \cite{10818605}. To overcome these limitations and to support cognitive/adaptive array behavior \cite{ref:s2022a, H12, G20}, we introduce a pre-steering strategy that enables a single network trained only at broadside to efficiently generalize to arbitrary desired source angles.
 
Pre-steering is a classical technique in array processing \cite{ref:capon1408a, ref:frost1972a, ref:friedlander1988a, ref:barabel_1984} that electronically aligns the array’s sensitivity with a desired look direction prior to beamforming. By applying initial phase shifts or time delays to the received signals, the array effectively perceives the desired source as arriving from broadside, simplifying adaptive processing and improving interference suppression. However, practical implementations are often imperfect due to calibration errors, quantization, or mismatch between assumed and actual source directions, which can degrade beamforming performance. Building on this insight, we extend pre-steering into a deep learning framework, training the network to recognize optimal sparse array configurations under both ideal and slightly perturbed steering conditions, thereby ensuring robust performance across a wide range of source directions and interference scenarios.

In this work, we implement a deep learning–based pre-steering strategy, where a single neural network is trained at broadside to predict optimal sparse array configurations, and pre-steered covariance inputs allow the network to generalize effectively to arbitrary desired directions. The pre-steered covariance is processed by the trained network to obtain the optimum sparse array configuration, which is then mapped back to the original angular domain for beamforming. By systematically introducing small, structured pre-steering perturbations during training, the network learns to tolerate angular mismatches and maintain near-optimal performance even when the actual source direction deviates slightly from the assumed look direction. This strategy leverages the rotational symmetry of the array manifold to generalize the network’s learning from broadside to any arbitrary direction, significantly reducing training complexity and dataset size while preserving high SINR performance across a wide range of source and interference angles. The approach also enhances robustness to practical errors in DOA estimation, making the framework well-suited for dynamic and uncertain propagation environments. Simulation results demonstrate that this DL-based pre-steering framework achieves over 90$\%$ accuracy in predicting optimal sparse array configurations, even under significant pre-steering errors, confirming both its efficiency and robustness for real-time adaptive beamforming.

\medskip
\noindent
\textbf{Contributions and Significance.}
The main contributions of this work are as follows:
\begin{itemize}
    \item We introduce a deep learning–based pre-steering framework that enables a single neural network trained at broadside to generalize to arbitrary desired directions, reducing training complexity and improving scalability.

    \item We investigate the robustness of the pre-steering method under steering angle mismatches and demonstrate that the approach maintains high accuracy under angular uncertainty.
    
    \item We show that the proposed method efficiently selects $M$ active antennas from $N$ uniformly spaced elements to maximize the output SINR for one desired source and one interference, enabling real-time sparse array adaptation.
\end{itemize}

The proposed framework provides a practical pathway toward real-time beamforming in dynamic environments, combining the efficiency of pre-steering with the inference capability of neural networks. The results demonstrate that the DL-based pre-steering strategy achieves near-optimal sparse array configurations with significantly reduced training overhead and enhanced robustness.

This paper is organized as follows. Section~\ref{sec:optimum_beamforming} deals with the problem formulation and receive array  data model.  Section~\ref{sec:minimum_projection_design} explains the pre steering concept. Section~\ref{Simulations} presents simulation examples using enumeration and DL methods, which is followed by conclusions in Section~\ref{sec:conclusion}.

\section{Optimum Sparse Array Beamforming} 
\label{sec:optimum_beamforming}
Consider  a desired source, which can be an emitter or a target,  and $L$ independent interfering sources whose signals impinge on a uniform linear array (ULA) with $N$ antennas. The baseband data received at the array at sampling instance $k$ is  given by, 
\begin {equation} \label{a}
\mathbf{x}(k)=   \alpha(k) \mathbf{a}( \theta_s)  + \sum_{i=1}^{L} \beta _i(k) \mathbf{a}( \theta_i)  + \mathbf{n}(k), 
\end {equation}
where ($\alpha (k)$, $\beta _i(k))$ $\in \mathbb{C}$  are the complex amplitudes of the incoming baseband signals, $\mathbf{n}(k)$ $\in \mathbb{C}^N$ is additive Gaussian noise  with  variance   $\sigma_n^2$, and ($\mathbf{a} ({\theta_s})$, $\mathbf{a} ({\theta_i})$) $\in \mathbb{C}^N$ are  the  respective steering vectors corresponding to the directions of arrival, $\theta_s$ and $\theta_i$, of the desired source and $i$th interference, and are defined as, 
\begin {eqnarray}  \label{b}
\mathbf{a} ({\theta_s})&=&[1 \,  \, e^{j (2 \pi / \lambda) d \text{sin}(\theta)  } \,  ...\ e^{j (2 \pi / \lambda) d (N-1) \text{sin}(\theta)  }]^\text{T} \nonumber \\
\mathbf{a} ({\theta_i})&=&[1 \,  \,  e^{j (2 \pi / \lambda) d \text{sin}(\theta_i)  } \,  ... \ e^{j (2 \pi / \lambda) d (N-1) \text{sin}(\theta_i)  }]^\text{T}
\end {eqnarray}
Here, $d$ is the inter-element spacing and the superscript `$\text{T}$' denotes matrix transpose.   
The received data $\mathbf{x}(k)$  is combined linearly by the $N$-sensor beamformer that strives to maximize the output SINR. The output signal $y(k)$ of the optimum beamformer for MaxSINR is given by \cite{1223538}, 
\begin {equation}  \label{c}
y(k) = \mathbf{w}_o^H \mathbf{x}(k), 
\end {equation} 
where the superscript `$H$' denotes Hermitian operation and $\mathbf{w}_o$ is the optimum weight vector resulting in the optimum output SINR$_o$. For statistically independent signals, the desired source correlation matrix is $\mathbf{R}_s= \sigma_s^2 \mathbf{a}( \theta_s)\mathbf{a}^H( \theta_s)$, where $ \sigma_s^2 =E\{\alpha (k)\alpha ^H(k)\}$. Likewise,  the  interference and noise correlation matrix, $\mathbf{R}_{s^{'}}= \sum_{i=1}^{L} (\sigma^2_i \mathbf{a}( \theta_i)\mathbf{a}^H( \theta_i)$) + $\sigma_n^2\mathbf{I}_{N\times N}$, with $ \sigma^2_i =E\{\beta _i(k)\beta_i^H(k)\}$ being the power of the $l$th interfering source.  There exists a  closed form solution of the optimum beamformer  and  is given by $\mathbf{w}_o = \mathscr{P} \{  \mathbf{R}_{s^{'}}^{-1} \mathbf{R}_s  \}$. The operator $\mathscr{P} \{. \}$  computes the principal eigenvector of the input matrix, yielding   the corresponding optimum output SINR$_o$;
\begin{equation}  \label{c2}
\text{SINR}_o=\frac {\mathbf{w}_o^H \mathbf{R}_s \mathbf{w}_o} { \mathbf{w}_o^H \mathbf{R}_{s^{'}} \mathbf{w}_o} = \Lambda_{max}\{\mathbf{R}^{-1}_{s^{'}} \mathbf{R}_s\}.
\end{equation}
This shows that the output SINR$_o$ is given by the maximum eigenvalue ($\Lambda_{max}$) associated with the product of  the inverse of interference plus noise correlation matrix  and the desired source correlation matrix. As indicated in  \cite{ref:hamza_sig_processing_2019, ref:wang2018a} it is clear that both the array weight vector and array configuration affect SINR, and their optimizations should optimally be pursued simultaneously. In this work, we aim to train a neural network to learn the sparse array configuration independently of the beamforming weights \cite{ref:hamza_wcnc_2024, 10548736}. Once the optimal array configuration is identified through network training, the corresponding beamformer weights are computed as a post-processing step. This formulation enables a focused evaluation of the network’s capability to capture the spatial structure and selectivity inherent in sparse array geometries, without the confounding influence of adaptive weight optimization.

\section{Pre-Steering} \label{sec:minimum_projection_design}
\subsection{Pre-Steering Approach}

The proposed \textit{pre-steering approach} exploits the rotational symmetry of the array manifold to enable a neural network trained only at broadside to generalize to arbitrary desired source directions. In contrast to conventional training procedures, where the network must learn mappings for all possible source angles, the pre-steering strategy aligns the desired signal direction with the broadside reference prior to inference. This is achieved by phase-rotating the received spatial covariance matrix using the steering vector corresponding to the desired direction. The transformation effectively converts the problem into an equivalent broadside scenario, allowing the trained network to infer the optimum sparse array configuration without additional retraining for new source angles. Once the configuration is obtained, it is mapped back to the original angular domain for beamforming. This process substantially reduces the training complexity, data storage, and generalization burden of the network while maintaining high accuracy across a wide range of angular conditions. Furthermore, by decoupling the training from explicit angle dependence, the proposed approach enhances robustness against steering angle mismatch and improves adaptability in dynamic propagation environments.

Let $\mathbf{R}_{x} \in \mathbb{C}^{N \times N}$ denote the spatial covariance matrix of the received signal, which includes contributions from the desired source, interference, and noise:
\begin{equation}
    \mathbf{R}_{x} = \sigma_{s}^{2} \mathbf{a}(\theta_{s}) \mathbf{a}^{H}(\theta_{s}) + \sigma_{i}^{2} \mathbf{a}(\theta_{i}) \mathbf{a}^{H}(\theta_{i}) + \sigma_{n}^{2} \mathbf{I}_{N\times N}
\end{equation}
where $\sigma_{s}^{2}$, $\sigma_{i}^{2}$, and $\sigma_{n}^{2}$ denote the source, interference, and noise powers, respectively.

To apply pre-steering, the received data are phase-rotated such that the desired direction $\theta_{s}$ is transformed to broadside. This can be expressed as
\begin{equation}
    \mathbf{R}_{\text{pre}} = \mathbf{D}(\theta_{s}) \, \mathbf{R}_{x} \, \mathbf{D}^{H}(\theta_{s}),
\end{equation}
where $\mathbf{D}(\theta_{s}) = \mathrm{diag}\{\mathbf{a}^{H}(\theta_{s})\}$ is a diagonal phase-shifting matrix that aligns the desired signal with the broadside axis.

The pre-steered covariance matrix $\mathbf{R}_{\text{pre}}$ is then provided as input to the trained neural network, which outputs the optimal sparse array configuration. The final beamforming weights $\mathbf{w}$ are subsequently computed using  MaxSINR formulations applied to the selected configuration. It is noted that  sparse array design can only have a few active sensors at a time which makes it difficult to  obtain full data  correlation values corresponding to the inactive sensor locations as required by the DL approaches. However, for the scope of this paper, we  assume that  estimates of all the  correlation lags corresponding to the  full  aperture array are available. This can typically be achieved by employing  a  low rank matrix completion  strategy or to sequentially estimate the missing data correlation values over different subarrays which are configured through antenna switching.

Through the pre-steering transformation, the network learns only a single angular representation (broadside) during training, yet remains applicable to any arbitrary desired direction during testing. This not only accelerates training and reduces the required dataset size but also enables fast and reliable array adaptation in real-time beamforming applications.

\section{Simulation Results} 
\label{Simulations} 
\subsection{Simulation Parameters}
The learning model is trained and tested on datasets simulated in MATLAB. 
We consider a  scenario with $N=12$ antennas and $M=6$ RF chains, where the antenna locations are evenly spaced with a half-wavelength inter-element spacing. Selecting 6 antennas out of the 12 available antenna locations, we obtain a total of 924 possible antenna configurations. The data set is divided into samples that each recorded the optimal sparse array for the given source and interference angle. For generating the  training datasets, the source angle remains fixed, while the interference angle iterates by a small angle, depending on the angle range and the number of training samples, across the angle range assumed in the field of view. The boundaries of this range and the number of samples are changed across several simulations to observe the effects. In each case, the SNR is fixed at 0 dB while the interference to noise ratio (INR) is varied between 15 and 20 dB to increase data diversity. The CNN testing datasets are similarly simulated by randomly selecting the interference angle within a given angle range and recording the optimal sparse array configuration. In some datasets, the source angle is fixed in the testing data, while in others it is also randomly selected within an angle range.

\subsection{Machine Learning Network Topology}
The employed CNN has an input  dimensions of size $3$×$N$×$N$ and comprises the real, imaginary parts, and phase of the $N$×$N$ correlation matrix. The output layer size depends on the number of unique sector configurations which varies depending on the training dataset. The neural network architecture implements 32 parallel filters, in each layer, of size 3×3 and 7×7 and incorporating dropout and batch normalization techniques. The network output is a one-hot encoded vector. The ReLU activation function is used for all layers except the output layer, which uses a softmax function. 
The network training utilized a batch size of 50 with a binary cross entropy loss function. The Adam optimizer \cite{ref:adam} is used with a cosine decay learning rate using an initial learning rate of $1^{-4}$, 10000 decay steps, and an alpha of $1^{-5}$. Up to 100 epochs are used for training, with an early stopping parameter set on the validation accuracy of the network and a patience of 25. 

\subsection{Sector Design}
\begin{figure}[]
    \centering
    \includegraphics[width=0.75\linewidth]{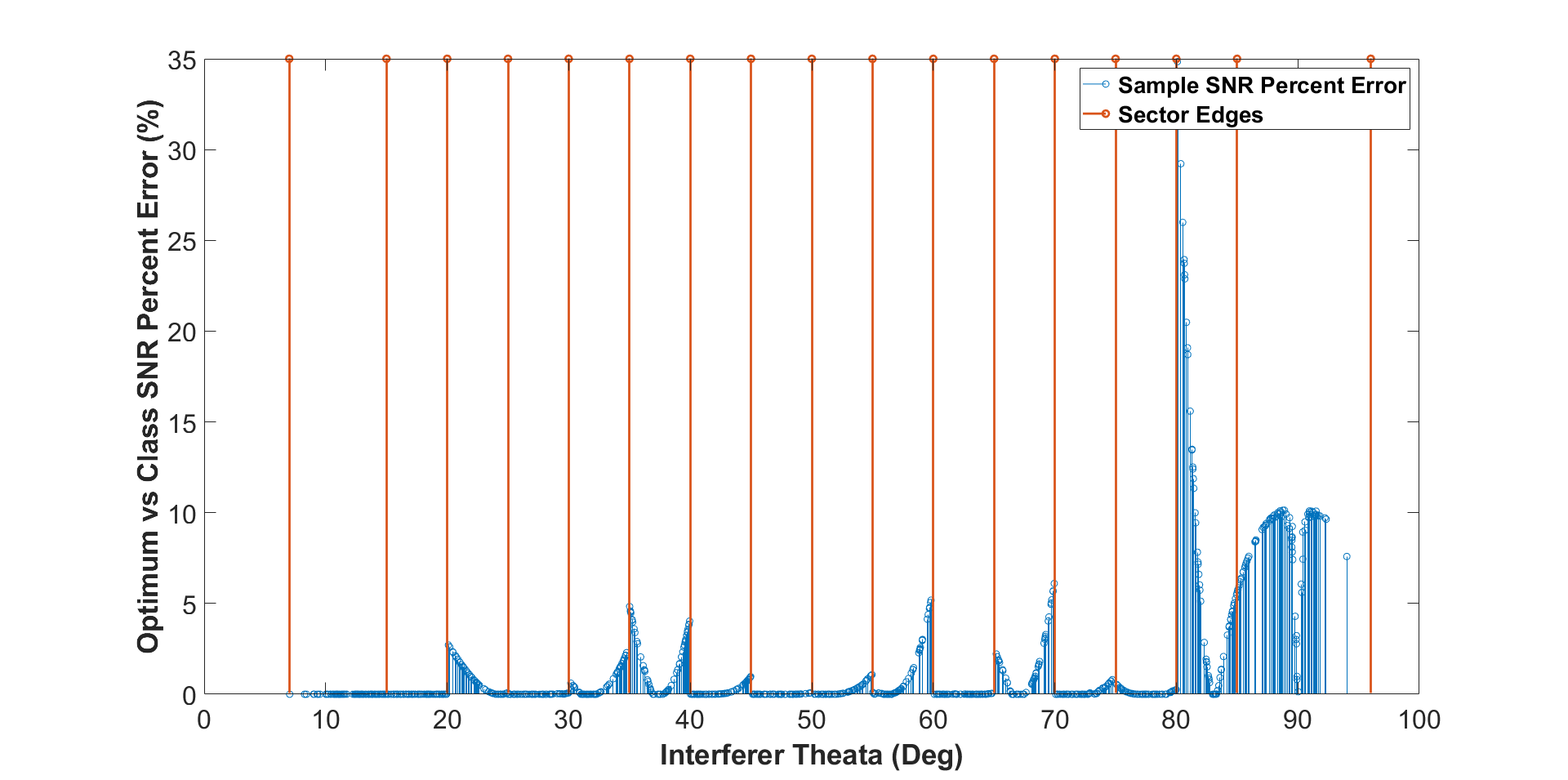}
    \caption{Sector array vs optimal array SINR percent error over interference angle range with 5 degree sectors.}
    \label{fig:five degree bins}
\end{figure}

One of the primary challenges in training a neural network to predict the configuration of a sparse antenna array lies in the combinatorial growth of possible configurations. For a system with $N=12$ possible antenna locations and $M=6$ RF chains, there are ($12$ choose $6$)$=924$ possible array configurations. This large number of discrete configurations poses a significant classification challenge for CNNs, as the number of output classes directly impacts both training complexity and accuracy.
To mitigate this issue, the field of view (FOV) is partitioned into angular bins referred to as sectors. Each sector is assigned the most frequently occurring optimal array configuration among the training interference angles samples within that angular range, rather than assigning an optimal array configuration for each interference angle. Consequently, the maximum number of CNN classes is limited by the number of sectors. In practice, multiple sectors often shared the same optimal configuration, further reducing the total number of unique classes.
However, this sector-based approach may degrade array performance in terms of SINR, since the configuration selected for a sector may not be individually optimal for every sample within that sector. Fig.~\ref{fig:five degree bins} illustrates the relationship between the optimal array SINR and the sector-assigned array SINR percentage error as a function of interference angle, ranging from 7$^{\circ}$ to 96$^{\circ}$. Each discrete blue marker represents a training data sample with a unique interference angle, while the orange vertical lines indicate the sector boundaries. For each sample, the SINR obtained using its optimal configuration is compared against the SINR achieved using the most common configuration in the corresponding sector, yielding the percentage error.

Overall, the percentage error remains low across most angles, demonstrating that the sector-based approximation provides acceptable performance. However, certain angles,  most notably around 80$^{\circ}$, exhibit significant deviations, indicating cases where the sector configuration does not align well with the true optimum. The error spike near 80$^{\circ}$ in Fig.~\ref{fig:five degree bins} arises because the bin lies close to broadside, where a single array configuration dominates within the limited set of 924 possibilities and, when applied to the entire bin, overrides configurations that are more optimal for angles near 80$^{\circ}$, leading to increased error.

\begin{figure}[]
    \centering
    \includegraphics[width=0.75\linewidth]{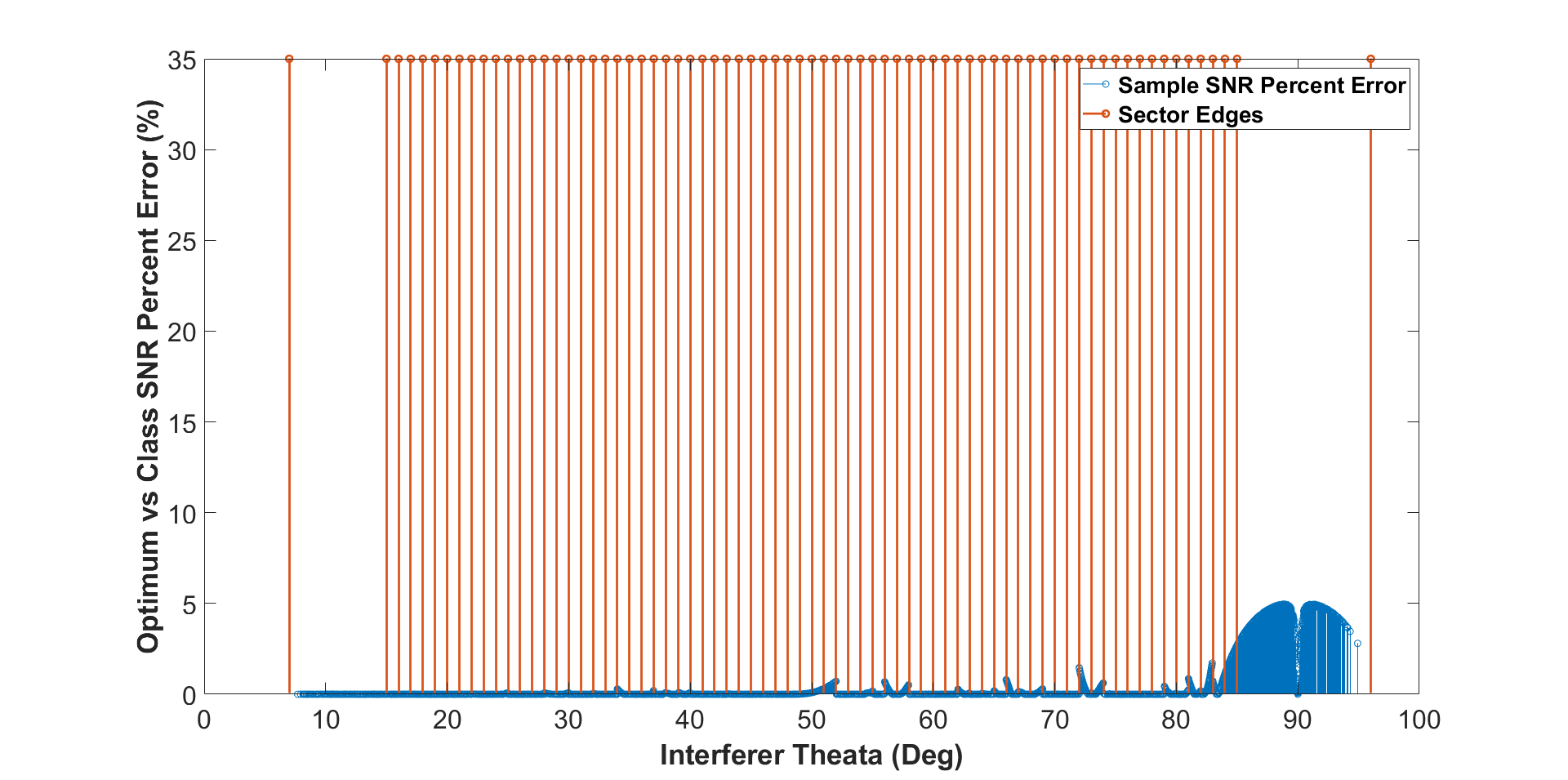}
    \caption{Sector array vs optimal array SINR percent error over interference angle range with 1 degree sectors.}
    \label{fig:one degree bins}
\end{figure}

\begin{figure}[]
    \centering
    \includegraphics[width=0.75\linewidth]{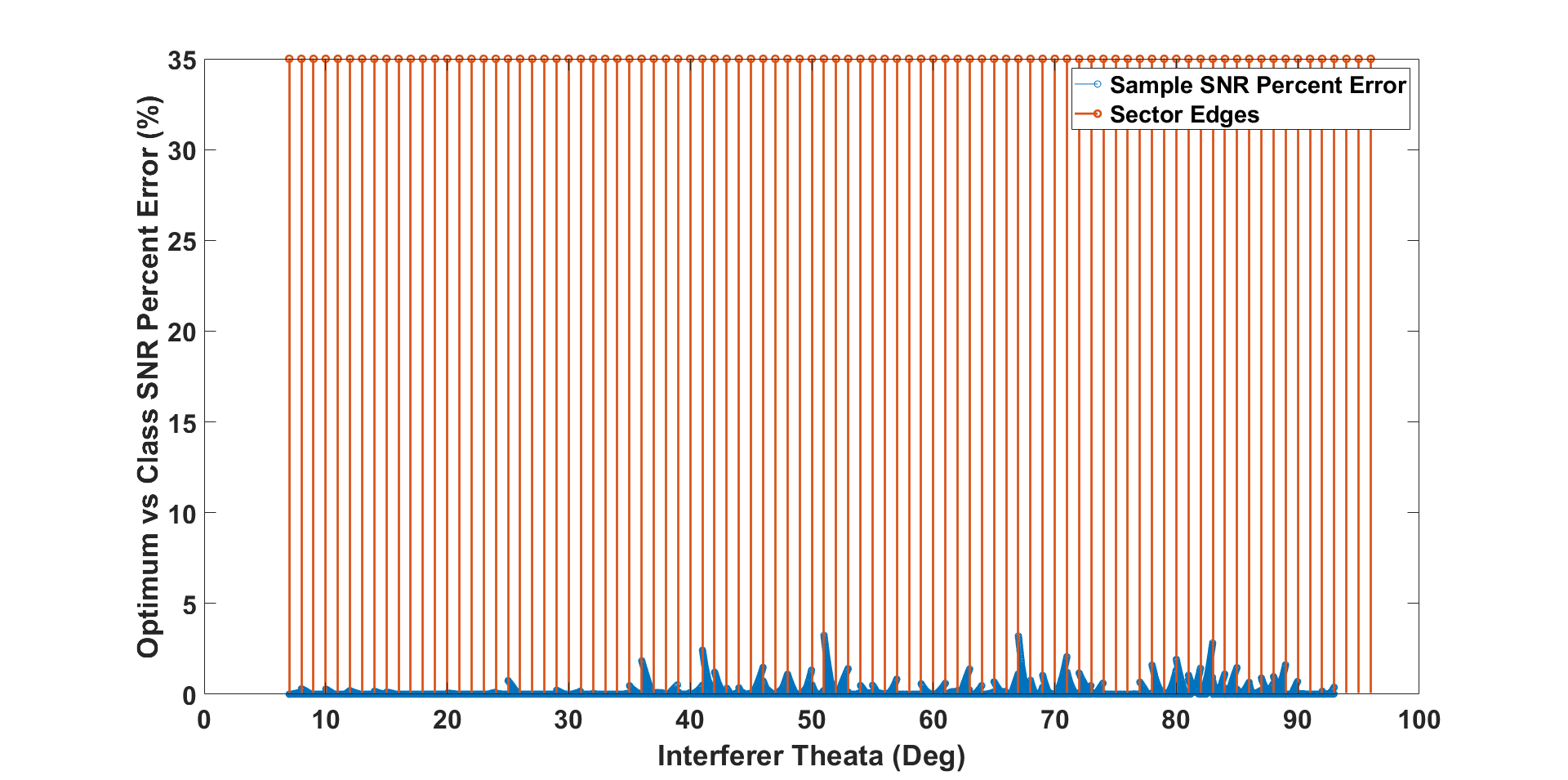}
    \caption{Sector array vs optimal array SINR percent error over interference angle range from three datasets with source angles at 30, 60, and 90 combined.}
    \label{fig:multi source}
\end{figure}

Reducing the angular span of each sector improves both the mean and maximum SINR error but increases the number of sectors, thus expanding the number of CNN output classes. Fig.~\ref{fig:one degree bins} presents the same analysis using one-degree sectors, showing reduced error variance at the cost of increased model complexity.
Using this dataset, the trained CNN achieves a testing accuracy of 94.38\%. It is important to note that this model is limited to predicting array configurations for a fixed source angle of 90$^{\circ}$, with interference angles ranging from 7$^{\circ}$ to 96$^{\circ}$.
To expand the operational range, additional datasets are generated for source angles of 30$^{\circ}$, 60$^{\circ}$, and 90$^{\circ}$, each comprising 8,000 samples. The combined dataset contained a total of 24,000 samples. Since the optimal array configuration varies significantly across source angles, sector assignments are recalculated independently for each source angle case, resulting in a total of 95 distinct CNN classes. The expanded dataset yields a testing accuracy of 49.96\%, with the corresponding SINR percentage error plotted in Fig.~\ref{fig:multi source}.
These results indicate that while the sector-based approach effectively reduces class complexity and maintains high accuracy for a single source angle, the method does not generalize well when extended across multiple source angles. The substantial reduction in testing accuracy highlights the sensitivity of the optimal array configuration to source direction and suggests that alternative modeling or feature representation strategies may be required for broader generalization.


\subsection{ Pre-steering design with robustness}
\label{sec:data_gen}
\begin{figure}[]
    \centering
    \includegraphics[width=0.75\linewidth]{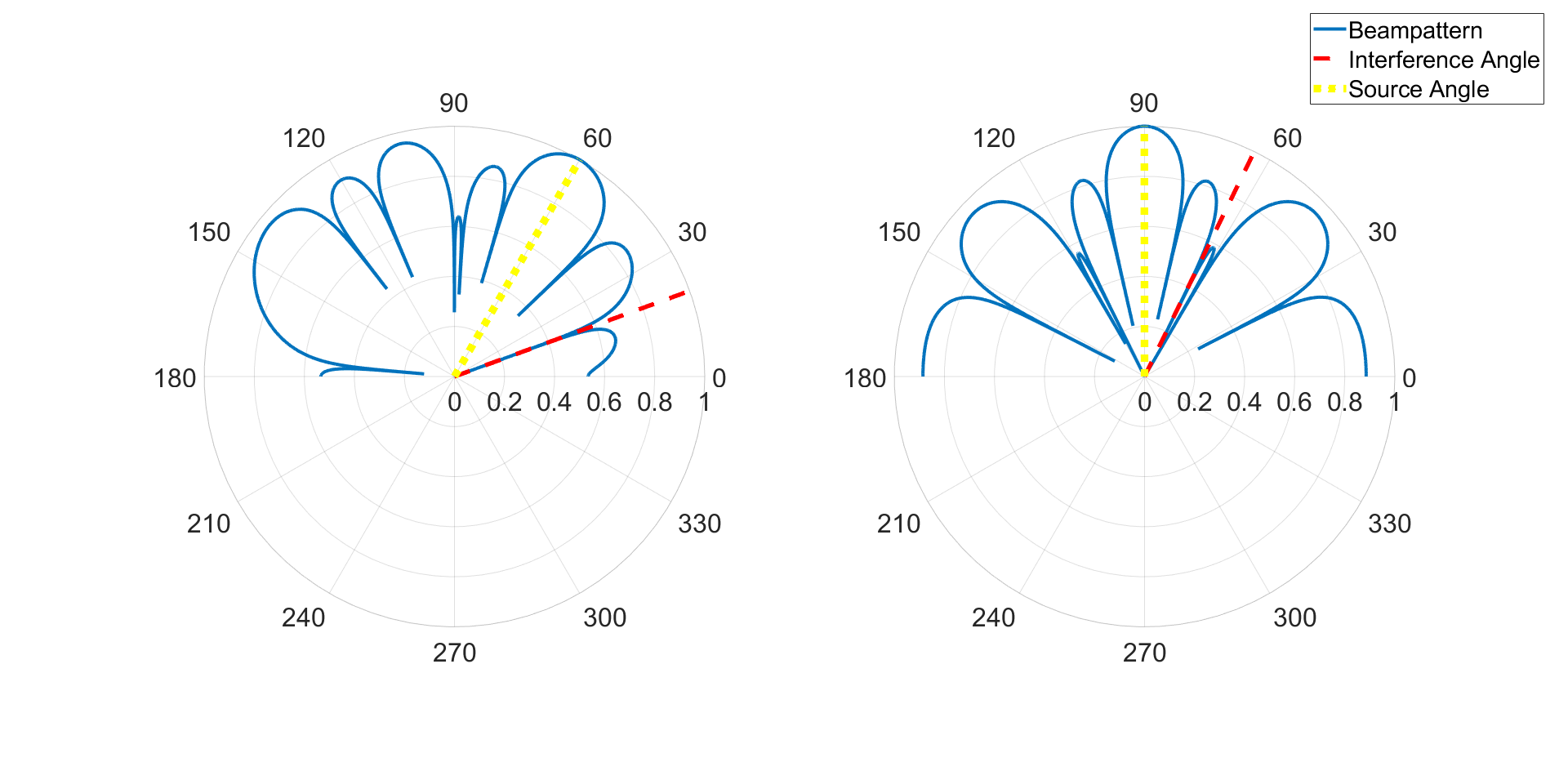}
    \caption{Beampattern, source, and interference angles before and after pre-steering shift.}
    \label{fig:pre-steer beampattern1}
\end{figure}

To evaluate the proposed pre-steering method, the training dataset is generated with a fixed source angle of 90$^{\circ}$ (broadside), while the interference angle varies from 7$^{\circ}$ to 173$^{\circ}$. The testing dataset is constructed with source angles ranging from 7$^{\circ}$ to 87$^{\circ}$ and interference angles from 10$^{\circ}$ to 87$^{\circ}$. For each test case, the resulting correlation matrix is pre-steered such that the source angle is shifted to 90$^{\circ}$, and the interference angle is transformed to a corresponding new value. As shown in Fig.~\ref{fig:pre-steer beampattern1}, the pre-steering operation realigns the source to broadside and maps the interference to its corresponding shifted angle, as reflected in the resulting beampattern. 


\begin{table}[h]
\label{tab:pre-steer solutions}
\renewcommand{\arraystretch}{1.2}
\resizebox{1\columnwidth}{!}{
\begin{tabular}{|c|c|c|c|}
\hline
\textbf{} & \textbf{Source angle ($^{\circ}$)} & \textbf{Interference angle ($^{\circ}$)} & \textbf{Optimal Solutions Indices} \\ \hline
\textbf{Before steer} & 60 & 20 & 605,883,233,796,754 \\ \hline
\textbf{After steer} & 90  & 63.9157 & 233,883,796,605,864 \\ \hline
\textbf{After steer with error} & 90.25 & 64.1937 & 233,605,883,796,524 \\ \hline
\end{tabular}
}
\caption{Summarizes that optimal sparse array solutions remain consistent after pre-steering even with pre-steering error, where the source angle does not shift exactly to broadside.}
\end{table}

As demonstrated in Table 1, the optimal array configurations remain consistent across the original, pre-steered, and error-perturbed cases; therefore, determining the optimal configuration from the pre-steered correlation matrix directly yields the optimal configuration for the original test scenario. Table 1 lists the greatest SINR array configurations, all of which can be considered optimal since their SINR performance differs by less than $10^{-14}$.
After pre-steering, all test samples exhibit source and interference angle relationships consistent with the training data (source at 90$^{\circ}$, interference between 7$^{\circ}$ and 173$^{\circ}$). This alignment enables the neural network to accurately categorize each test case and identify the corresponding optimal array configuration.

Using this approach, the proposed neural network achieves a classification accuracy of 91.84\%, demonstrating the effectiveness of the pre-steering strategy in improving generalization performance across varying source and interference cases. 


The proposed approach assumes ideal pre-steering without error; however, in practice, the pre-steering operation can introduce inaccuracies when the source angle is not accurately known. To evaluate robustness to imperfect knowledge of the desired source DOA, controlled perturbations are introduced to the pre-steering angle during both training and testing to assess whether the neural network can still predict the optimal antenna configuration under such errors.

In the initial experiment, the training data is generated twice, once without error and once with pre-steering error. The nominal source angle, $\theta s$, is fixed at 90$^{\circ}$, while the interference angle, $\theta i$, varies from 7$^{\circ}$ to 173$^{\circ}$. For the error-added dataset, a small random perturbation of -0.2$^{\circ}$, 0$^{\circ}$, or +0.2$^{\circ}$ is added to $\theta s$ for each sample to simulate pre-steering inaccuracies.

The same approach is applied to the testing dataset. During testing, $\theta s$ values are randomly sampled between 7$^{\circ}$ and 87$^{\circ}$and $\theta i$ values are sampled between 10$^{\circ}$ and 87$^{\circ}$. Each sample is pre-steered to make $\theta s$=90$^{\circ}$, but random pre-steering errors (-0.2$^{\circ}$, 0$^{\circ}$, +0.2$^{\circ}$) are again added to simulate misalignment. Using this data generation method, the trained neural network achieves a testing accuracy of 25.62\% shown in the first row of Table. 2, indicating limited robustness under pre-steering error.

The random perturbations introduced inconsistent optimal configurations, making learning difficult. To address this, we generated multiple samples for each configuration, all sharing the same optimal antenna configuration but with slightly different pre-steering errors. For each interference angle $\theta i$, five samples are generated corresponding to pre-steering errors of -0.5$^{\circ}$, -0.25$^{\circ}$, 0$^{\circ}$, +0.25$^{\circ}$, and +0.5$^{\circ}$, respectively. The optimal configuration is computed only for the error-free case (0$^{\circ}$). This modified dataset improves the testing accuracy to 33.03\% shown in the second row of Table. 2, demonstrating that structured pre-steering error sampling provided slightly better generalization.

Next, the pre-steering error range is reduced to -0.2$^{\circ}$, -0.1$^{\circ}$, 0$^{\circ}$, +0.1$^{\circ}$, and +0.2$^{\circ}$ to minimize excessive variance while maintaining robustness.  Even when tested with a larger random error of ±0.25$^{\circ}$ ±0.25$^{\circ}$, the network achieves a significantly improved accuracy of 57.63\% shown in the third row of Table. 2, suggesting that the smaller training error range enhances the model’s ability to generalize to unseen pre-steering perturbations.

To further improve robustness, the size of the training dataset is doubled from 8000 to 16000 samples. The source angle $\theta s$ is fixed at 90$^{\circ}$, and $\theta i$ varied from 7$^{\circ}$ to 173$^{\circ}$ in increments of approximately 0.05. Five samples are generated per $\theta i$ using pre-steering errors of -0.2$^{\circ}$, -0.1$^{\circ}$, 0$^{\circ}$, +0.1$^{\circ}$, and +0.2$^{\circ}$. In this configuration, the training set includes small, systematic pre-steering errors, while the testing set contained larger random errors (±1$^{\circ}$). Despite the increased test error magnitude, the network achieved a testing accuracy of 90.66\%, demonstrating strong robustness to pre-steering imperfections. Table. 2, summaries the network accuracy under the various pre-steering error strategies.


\begin{table}[h]
\label{tab:Methods Compared}
\renewcommand{\arraystretch}{1.2}
\resizebox{1\columnwidth}{!}{
\begin{tabular}{|c|c|c|c|}
\hline
\textbf{Experiment} & \textbf{Training Error Range ($^{\circ}$)} & \textbf{Testing Error Range ($^{\circ}$)} & \textbf{Accuracy (\%)} \\ \hline
\textbf{Random small error} & {-0.2, 0, +0.2} & {-0.2, 0, +0.2} & 25.62 \\ \hline
\textbf{Structured error samples} & {-0.5, -0.25, 0, +0.25, +0.5}           & {-0.5, 0, +0.5}             & 33.03 \\ \hline
\textbf{Reduced structured error} & {-0.2, -0.1, 0, +0.1, +0.2}           & {-0.25, 0, +0.25}             & 57.63 \\ \hline
\textbf{Expanded dataset} & {-0.2, -0.1, 0, +0.1, +0.2}           & {-1, 0, +1}             & 90.66 \\ \hline
\end{tabular}
}
\caption{Summarizes the network accuracy under various pre-steering error strategies discussed.}
\end{table}


\section{Conclusion}
\label{sec:conclusion}
The paper presented a machine learning–based approach for optimizing sparse antenna array configurations using CNNs. To address the large combinatorial space of possible array selections, sector-based labeling and correlation-matrix–based feature representations were employed to reduce model complexity while maintaining array performance. The results demonstrated that this strategy provides high classification accuracy for fixed source-angle scenarios and that sectorization offers an effective balance between tractability and SINR degradation. To extend the method to more practical settings in which the source angle may vary, a pre-steering transformation was introduced to map all test scenarios into a common angular reference frame. Under ideal pre-steering conditions, the neural network generalized well across wide source and interference angle ranges, achieving accuracies exceeding 90$\%$. However, this work also showed that even small pre-steering errors can significantly degrade performance if such imperfections are not represented in the training data.
To mitigate this vulnerability, structured pre-steering error augmentation was incorporated. By generating multiple training samples per interference angle with controlled perturbations, the network learned to tolerate imperfections in DOA alignment. Reducing the error range and expanding the training dataset further enhanced performance. The proposed approach achieved strong generalization capability, maintaining over 90$\%$ accuracy even under large, randomly generated pre-steering errors.
These findings highlight the importance of structured data diversity and error-aware training strategies for achieving robust sparse array configuration prediction.
 
\balance

\bibliographystyle{IEEEtran}
\bibliography{references.bib}
\end{document}